\documentclass{webofc}
\usepackage[varg]{txfonts}   
\usepackage{graphicx,colordvi}
\begin{document}
\title{Fission Fragment Mass and Total Kinetic Energy Distributions of 
  Spontaneously Fissioning Plutonium Isotopes}
\author{K. Pomorski, B. Nerlo-Pomorska, J. Bartel$^*$, C. Schmitt$^{*, **}$ }
\email{Krzysztof.Pomorski@umcs.pl}
\institute{UMCS, Lublin, Poland,{$^*$} IPHC, Strasbourg, France,
  {$^{**}$} GANIL, CAEN, France}
\abstract{The fission-fragment mass and total kinetic energy (TKE) distributions
are evaluated in a quantum mechanical framework using elongation, mass
asymmetry, neck degree of freedom as the relevant collective parameters in the
Fourier shape parametrization recently developed by us. The potential energy
surfaces (PES) are calculated within the macroscopic-microscopic model based on
the Lublin-Strasbourg Drop (LSD), the Yukawa-folded (YF) single-particle
potential and a monopole pairing force. The PES are presented and analysed in
detail for even-even Plutonium isotopes with $A=236 -246$. They reveal deep
asymmetric valleys. The fission-fragment mass and TKE distributions are obtained
from the ground state of a collective Hamiltonian computed within the
Born-Oppenheimer approximation, in the WKB approach by introducing a
neck-dependent fission probability. The calculated mass and total kinetic energy
distributions are found in good agreement with the data.}


\maketitle


\section{Introduction}

Our present understanding of nuclear fission is still based on the idea of
Lisa Meitner and Otto Frisch \cite{MF39} of a deformed charged liquid drop. In a seminal
paper \cite{BW39} Niels Bohr and John A. Wheeler fully developed the concept of
the energy surface of a nucleus as function of a set of deformation parameters.
The height of the minimal energy barrier that the nucleus has to overcome in the
multidimensional energy hyper-surface as function of the deformation parameters
on its decay path determines the stability of the nucleus against fission.
Unfortunately, following the Bohr-Wheeler paper, one sometimes still uses up to
now the Lord Rayleigh expansion of the nuclear surface into spherical harmonics.
It was shown (e.g. in Ref.~\cite{DPB07}) that such an expansion is not rapidly
converging for large deformations and one needs to include terms up to 
multipolarity 16 in order to
describe shapes close to the scission point. Fortunately, some better
parametrizations of nuclear shapes exist. Among the most popular are the
Quadratic Surfaces of Revolution (QSR) originally proposed by Nix \cite{Nix},
the Funny-Hills parametrization and its extension \cite{FH72,MFH}, and the
Cassini ovals with modifications due to Pashkevich \cite{Pash}. The first two
of the above quoted parametrizations are not analytical and closed (e.g. do no 
allow for the inclusion of higher order terms) while the last one is not easy to
handle and its parameters do not have a clear physical interpretation. 

The recently developed Fourier parametrization (see Appendix) of deformed
nuclear shapes \cite{PNB15,PRC17} is free of these ambiguities and is rapidly
converging. Using this para\-metrisation we have made an attempt to obtain the
fission-fragment mass and total kinetic energy distribution of even-even
Plutonium isotopes with mass numbers $236\le A\le 246$. The
deformation-energy landscapes that we are going to discuss in Section 4
are described by three Fourier deformation parameters $q_2,\,q_3,\, q_4$ which
are respectively elongation, mass-asymmetry and neck degree of freedom of
the fissioning nucleus \cite{PNB15,PRC17}. Nonaxial shapes were not included in
the present research since at large nuclear deformations they turn out to play
a minor role. Thanks the fast convergence of the Fourier series, we
believe that the above defined three deformation parameters are sufficient to
describe the large variety of nuclear shapes up to very large deformations
\cite{PRC17}.The potential energy surfaces of different fissioning nuclei were
calculated within the macroscopic-microscopic method \cite{FH72,BCS}. The 
fragment mass distribution obtained in low-energy fission of light actinides was
evaluated in a quantum mechanics framework within the Born-Oppenheimer
approximation (BOA) \cite{NPI15}. 

Within the aforementioned deformation space the fission yield is obtained from the probability
distribution of the collective wave function on the ($q_3,\,q_4$) plane in the 
vicinity of the scission configuration ($q_2 \approx 2.3$). A neck-size
dependent fission probability \cite{PNI16} was used to evaluate the
fission-fragment mass and kinetic energy yields from the distribution
probability at different elongations of the fissioning nucleus as will be
explained in Section 3.


\section{Macroscopic-microscopic model of the potential energy}

The nuclear deformation energies of our analysis were determined in the
macroscopic-microscopic approach, where the Lublin Strasbourg Drop (LSD) model
\cite{LSD} has been used for the macroscopic part of the potential-energy
surface. Microscopic effects have been evaluated through a Yukawa-folded (YF)
single-particle potential \cite{YF} with the parameters listed in
Ref.~\cite{DPB16}, where also our way of solving of the eigenproblem of the YF
Hamiltonian is described. Eighteen deformed harmonic oscillator shells were
taken into account when diagonalising the YF Hamiltonian. The Strutinsky
shell-correction method \cite{Str66,FH72} with a $8^{\rm th}$ order correctional
polynomial and a smearing width $\gamma_S=1.2\,\hbar\omega_0$ is used, where
$\hbar\omega_0=41/A^{1/3}$ MeV is the spherical harmonic-oscillator frequency.
The BCS \cite{BCS} theory, including an approximate GCM+GOA particle-number
projection as described in Refs.~\cite{APBCS} was used for the pairing
correlations. The pairing strength equal to $G\cdot {\cal
N}^{2/3}=0.28\,\hbar\omega_0$, (with ${\cal N}=Z,N$ for protons or neutrons) was
adjusted to the experimental mass differences of nuclei in this region using a
pairing window composed of $2\sqrt{15{\cal N}}$ single-particle levels closest
to the Fermi surface \cite{PPS89}.

\begin{figure*}[t!]
\centerline{
\includegraphics[width=0.4\textwidth,angle=0]{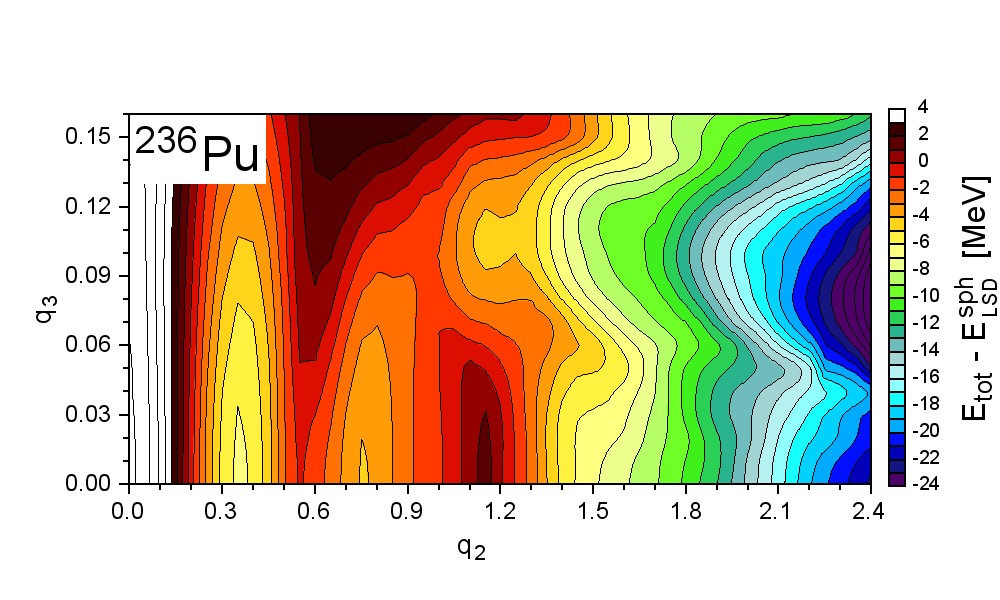}\hspace{1cm}
\includegraphics[width=0.4\textwidth,angle=0]{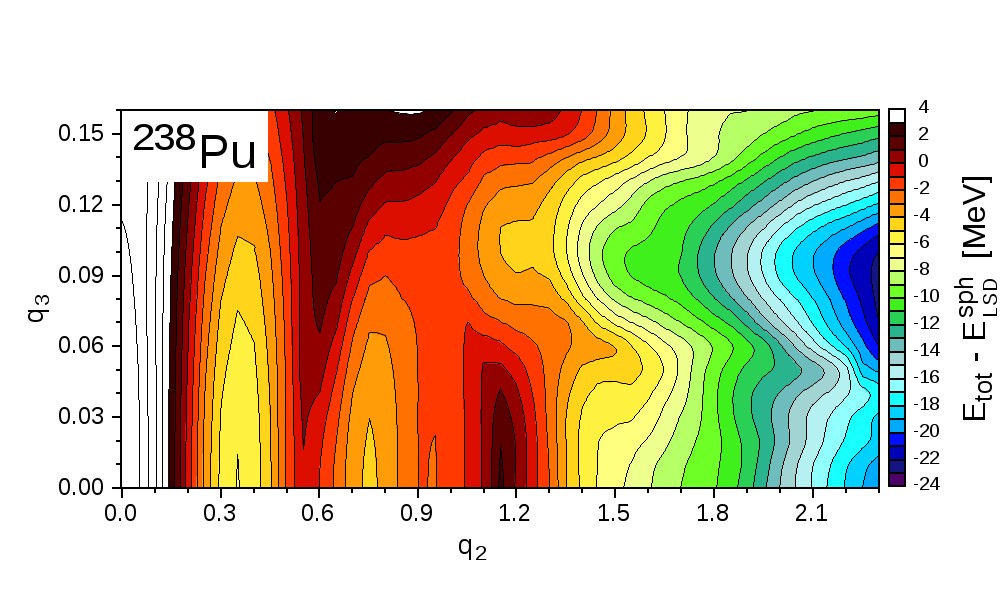}}
\centerline{
\includegraphics[width=0.4\textwidth,angle=0]{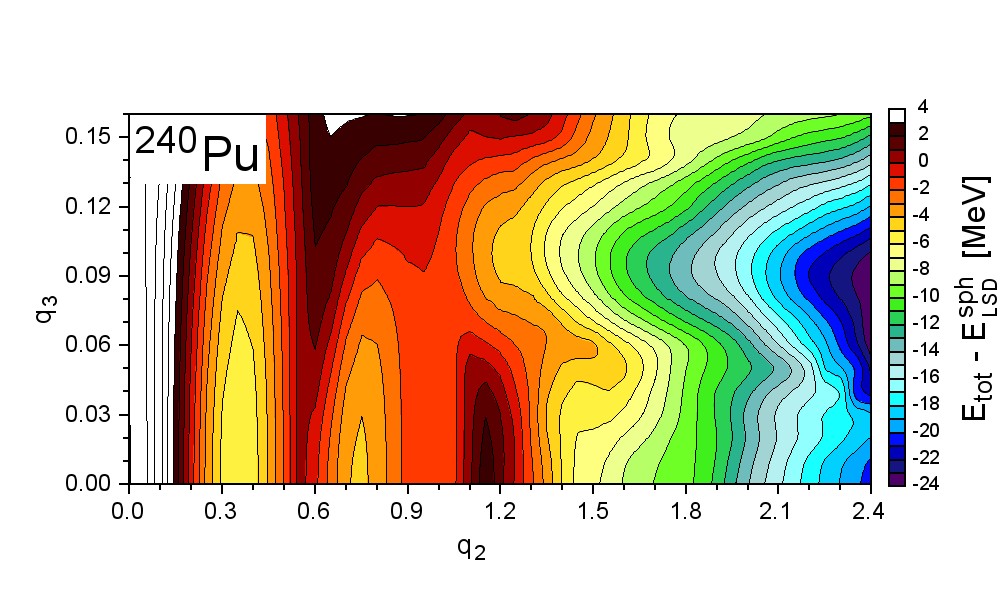}\hspace{1cm}
\includegraphics[width=0.4\textwidth,angle=0]{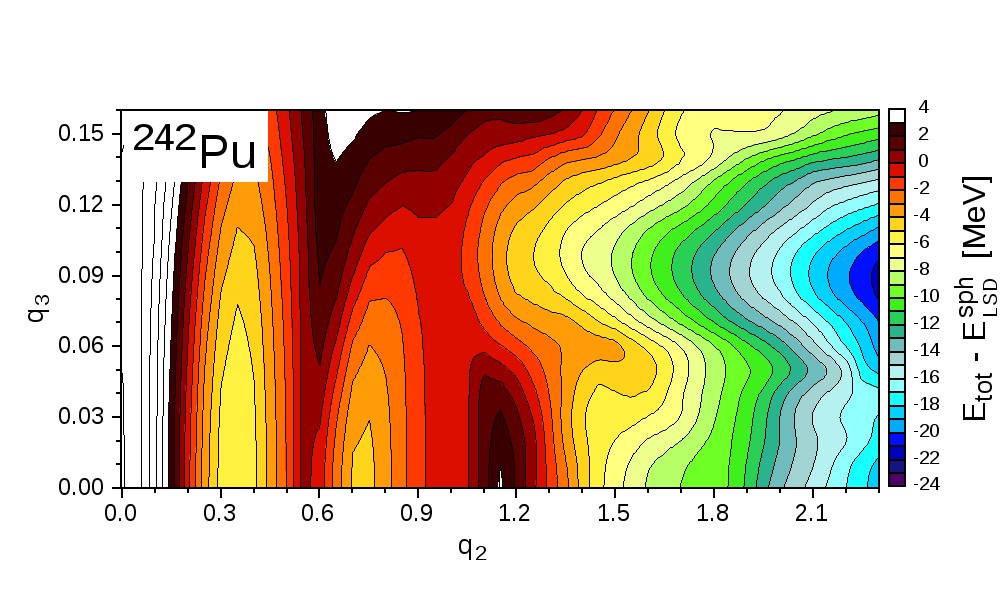}}
\centerline{
\includegraphics[width=0.4\textwidth,angle=0]{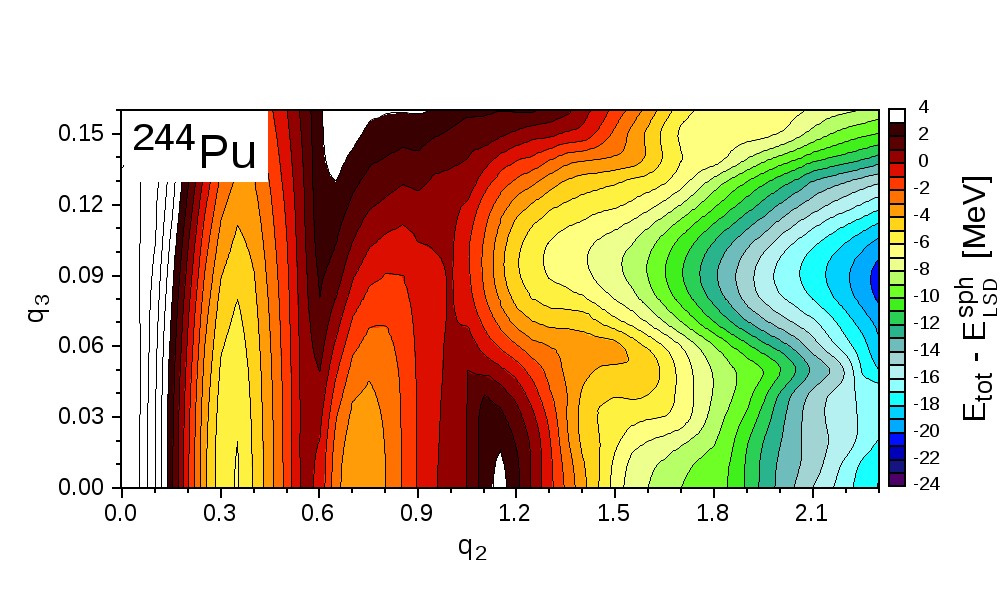}\hspace{1cm}
\includegraphics[width=0.4\textwidth,angle=0]{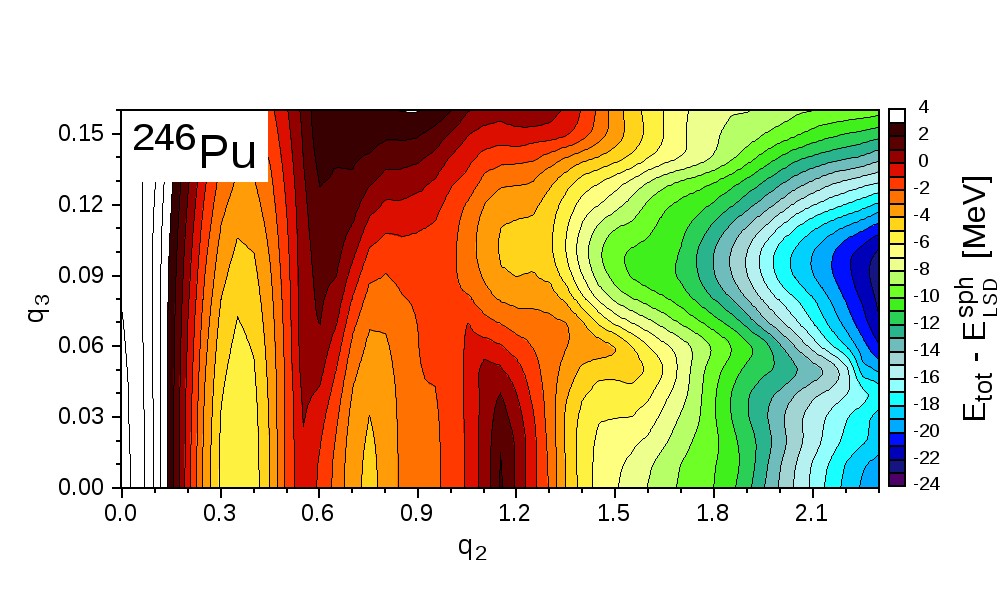}}
\caption{Potential energy surfaces on the $(q_2,q_3)$ plane
of even-even Pu isotopes minimized with respect to $q_4$.} 
\label{pus}
\end{figure*}

\section{Collective model of fission}

In any coordinates $q_n$ the collective Hamiltonian has the following 
form:
\begin{equation}
\widehat H_{\rm coll}=-\frac{\hbar^2}{2} \sum_{i,j} |M|^{-1/2}
  \frac{\partial}{\partial q_i}
 |M|^{-1/2} M^{-1}_{ij}\frac{\partial}{\partial q_j} + V~,
\end{equation}
where $M_{ij}(\{q_i\})$ and $V(\{q_i\})$ denote the inertia tensor and the
potential energy, respectively and $|M|=\det(M_{ij})$.

The eigenproblem of this Hamiltonian can be solved in the BOA in which one
assumes that the motion towards fission, i.e. in the $q_2$ coordinate, is much slower than the motion in the $q_3$ and $q_4$ collective variables. This implies that the eigenfunction of $\widehat H_{\rm coll}$ can be approximated in our 
case in the following form:
\begin{equation}
\Psi_{nE}(q_2,q_3,q_4)=
     u_{nE}(q_2) \; \varphi_{n}(q_3,q_4;q_2)~.
\end{equation}
Here $u_{nE}(q_2)$ is the wave function for the fission (elongation) mode and
$\varphi_n$ are the eigenfunctions of the Hamiltonian which describe the
collective motion in the perpendicular directions. In the following we shall use
the WKB approximation for the $u_{nE}(q_2)$ wave function and consider only the
lowest energy eigenstate $\varphi_{n=0}$ since we are interested in fission at
very low excitation energy. The effect of taking into account higher states was discussed in Ref.~\cite{NPI15}.

The probability of finding the system, for a given elongation $q_2$, in a 
deformation defined by ($q_3,q_4$) is equal to
\begin{equation}
W(q3,q4;q2) = |\Psi(q_3,q_4;q_2)|^2 = |\varphi_0(q_3,q_4;q_2)|^2
\label{dipo}
\end{equation}
Our model is still simplified further. Instead of the square of the 
collective wave function (\ref{dipo}), we take the following Wigner function:
\begin{equation}
W(q3,q4;q2) \sim \exp\left\{-\frac{V(q_3,q_4;q_2)-V_{\rm eq}(q_2)}
     {E_0+T}\right\}
\label{Wig}
\end{equation}
where $V_{\rm eq}(q_2)$ is the potential minimum for a given $q_2$,
$E_0$ is the zero-point energy, treated here as a free parameter and $T$ is the
temperature of the nucleus.

The probability distribution integrated over $q_4$
\begin{equation}
 w(q_3;q_2)=\int W(q_3,q_4;q_2)\, dq_4~,
\label{waR}
\end{equation}
is directly related to the fragment mass yield at given elongation $q_2$.

It is obvious that the fission probability should also depend on the neck
radius. Following Ref.~\cite{PNI16} we assume a neck-rupture probability $P$
in the form:
\begin{equation}
  P(q_2,q_3,q_4)=\frac{k_0}{k} P_{\rm neck}(R_{\rm neck})\,,
\label{pn}
\end{equation}
where $k$ is the momentum in the direction towards fission (or simply the 
velocity along $q_2$), while $R_{\rm neck}(q_2,q_3,q_4)$ is the 
deformation-dependent neck radius. $k_0$ plays the role of a scaling parameter.
The neck rupture probability \cite{PNI16} can be taken e.g. in one of the
following forms:
\begin{equation}
P_{\rm neck}(R_{\rm neck})=\left\{
\begin{array}{lr}
\exp[-\ln 2(R_{\rm neck}/d)^2]  & {\it Gauss}\\[1ex]
1/\cosh[(2+\sqrt{3})R_{\rm neck}/d] & {\it 1/cosh} \\[1ex]
1/[1+(R_{\rm neck}/d)^2]    & {\it Lorentz}
\end{array}
\right.
\label{pneck}
\end{equation}
The parameter $d$ is the {\it half-width} of the probability and is treated
here as a free adjustable parameter.

The momentum $k$ in Eq.~(\ref{pn}) has to ensure that the probability depends
on the time in which one crosses the subsequent interval in $q_2$: $\Delta t=
\Delta q_2/v(q_2)$, where 
\begin{equation}
v(q_2)=\hbar k/\overline M(q_2)
\label{velocity}
\end{equation}
is the velocity towards fission. The inertia $\overline M(q_2)$ is evaluated 
using the approximation proposed in Ref.~\cite{RLM76}:
\begin{equation}
\overline M(q_2) = \mu \left[1+11.5\cdot(B_{\rm irr}/\mu -1)\right]
     \left(\frac{\partial R_{12}}{\partial q_2}\right)^2~,
\end{equation}
where $B_{\rm irr}$ is the irrotational inertia corresponding to the distance
$R_{12}$ between the fragments and $\mu$ is the reduced mass. The value of $k$ 
in Eq.~(\ref{velocity}) depends on the difference $E-V(q_2)$ and on the part of
the collective energy which is converted into heat $Q$:
\begin{equation}\label{kk}
\frac{\hbar^2 k^2}{2\overline M(q_2)}=E_{kin}=E-Q - V(q_2)~.
\end{equation}

In order to introduce the neck-rupture probability $P$, in our formalism one has
to rewrite the probability distribution $w(q_3;q_2)$ in the form
\begin{equation}
 w(q_3;q_2)=\int W(q3,q4;q2) P(q_3,q_4,q_2)\,dq_4~.
\label{pda}
\end{equation}
Such an approximation means that the fission process is spread over some region
of $q_2$ and that for a given $q_2$, at fixed mass asymmetry, one has to take
into account the probability to fission at a previous $q_2$ value, i.e. one
has to replace $w(q_3;q_2)$ by
\begin{equation}
w^{\prime}(q_3;q_2)=w(q_3;q_2) \; \frac{1-\int\limits_{q^{\prime}_2\le q_2}
 \!\!\! w(q_3;q^{\prime}_2)\,d q^{\prime}_2}{
 \int\limits^{ } w(q_3;q^{\prime}_2)\,d q^{\prime}_2}~.
\end{equation}
The integral mass yield can then be written as the sum of all partial yields at
different $q_2$:
\begin{equation}
 Y(q_3)=\int\limits_{ } w^{\prime}(q_3; q_2)\,dq_2\,\,/\, 
   \int\limits^{ } w^{\prime}(q_3; q_2)\,d q_2\, dq_3~.
\label{Ya}
\end{equation}
As one notices from (\ref{Ya}), the scaling factor $k_0$ in the expression for 
$P$, Eq.~(\ref{pn}), has vanished and does no longer appear in the definition 
of the mass yield. Our model thus has only two adjustable parameters, 
$E_0$ in Eq.\ (\ref{Wig}) and the width parameter $d$, that appears in the 
neck-rupture probability (\ref{pneck}).
\\[-4ex]


\section{Results}
The potential energy surfaces (PES), relative to the corresponding 
spherical LSD energy, were calculated for even-even Plutonium isotopes 
$^{236-246}$Pu in the 3D $(q_2,q_3,q_4)$ deformation space,
 where $q_2$ describes the elongation of the nucleus, $q_3$ its
 reflection asymmetry, and $q_4$ controls the neck size [8,9].
The
$(q_2,q_3;q_4^{\rm min})$ cross-sec\-tions of the PES minimized with respect to
$q_4$ are presented in Fig.~\ref{pus}. Pronounced mass asymmetric $(q_3\ne 0)$
fission valleys are visible in all isotopes beyond $q_2 \approx 1$. Symmetric
fission valleys are generally present but lay always higher by a few MeV, and 
with a substantial ridge separating them from the asymmetric valleys.\\[-4ex]
\begin{figure}[h]
\centerline{\includegraphics[width=0.8\textwidth,angle=0]{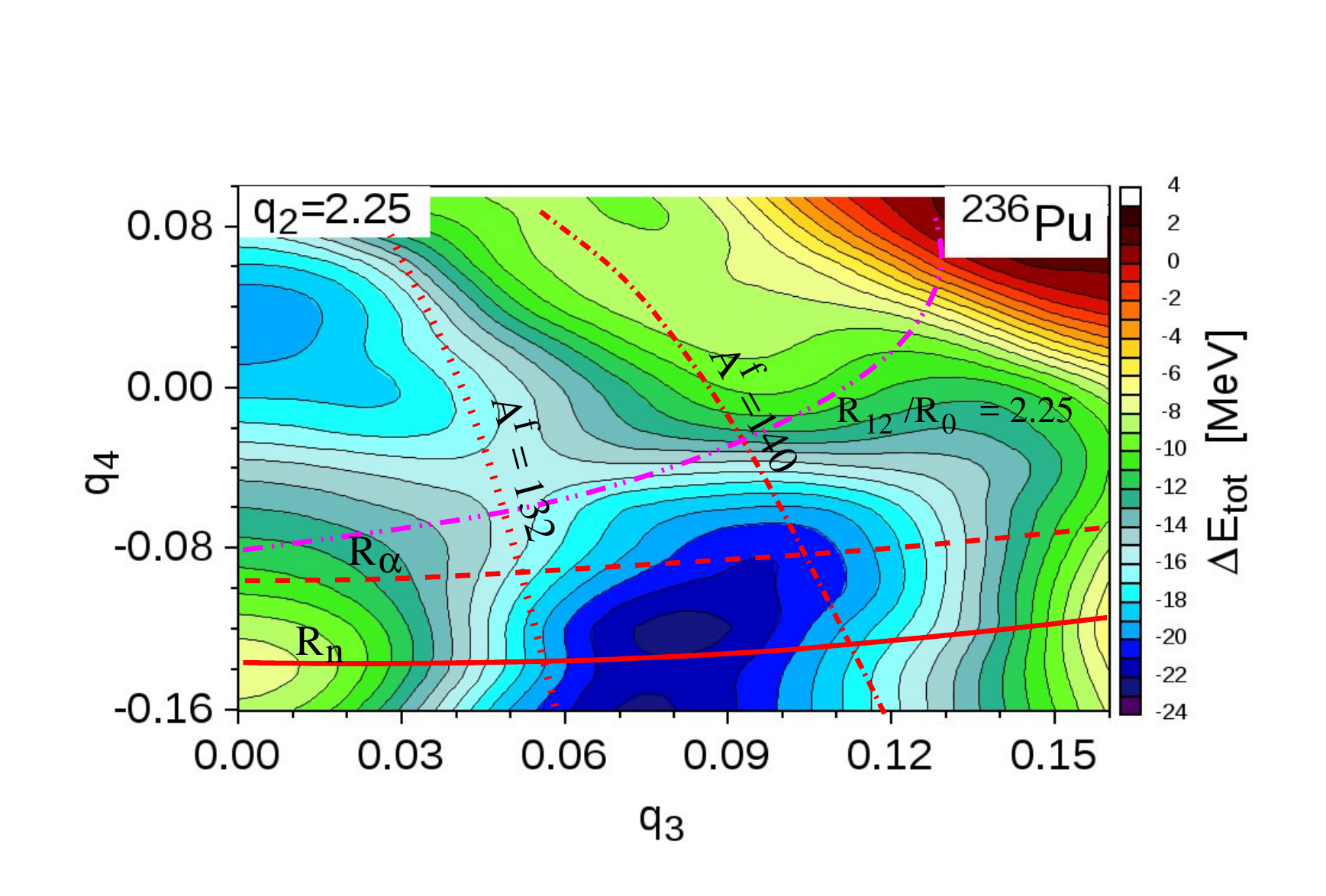}}
\caption{$(q_3,q_4)$ PES cross-section of $^{236}$Pu for $q_2=2.25$.}
\label{e34}
\end{figure}
\begin{figure}[thb!]
\centerline{\includegraphics[width=0.6\textwidth,angle=0]{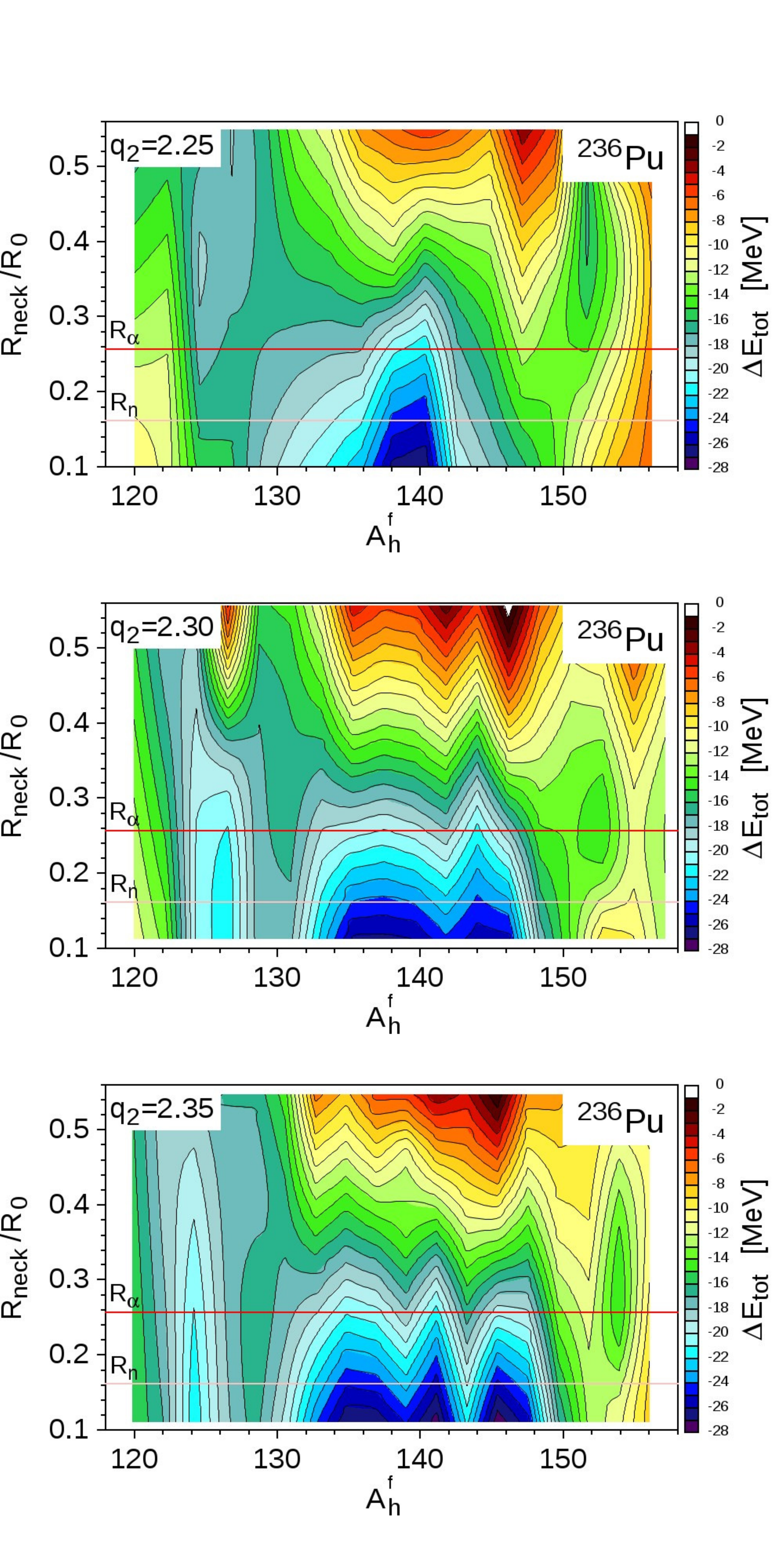}}
\vspace{-3mm}
\caption{Potential energy surfaces for different elongations $q_2$
   projected onto the $(A^f_h,R_{\rm neck})$ plane.} 
\label{erna}
\vspace{-10mm}
\end{figure}
\begin{figure}[thb]
\centerline{\includegraphics[width=0.6\textwidth,angle=0]{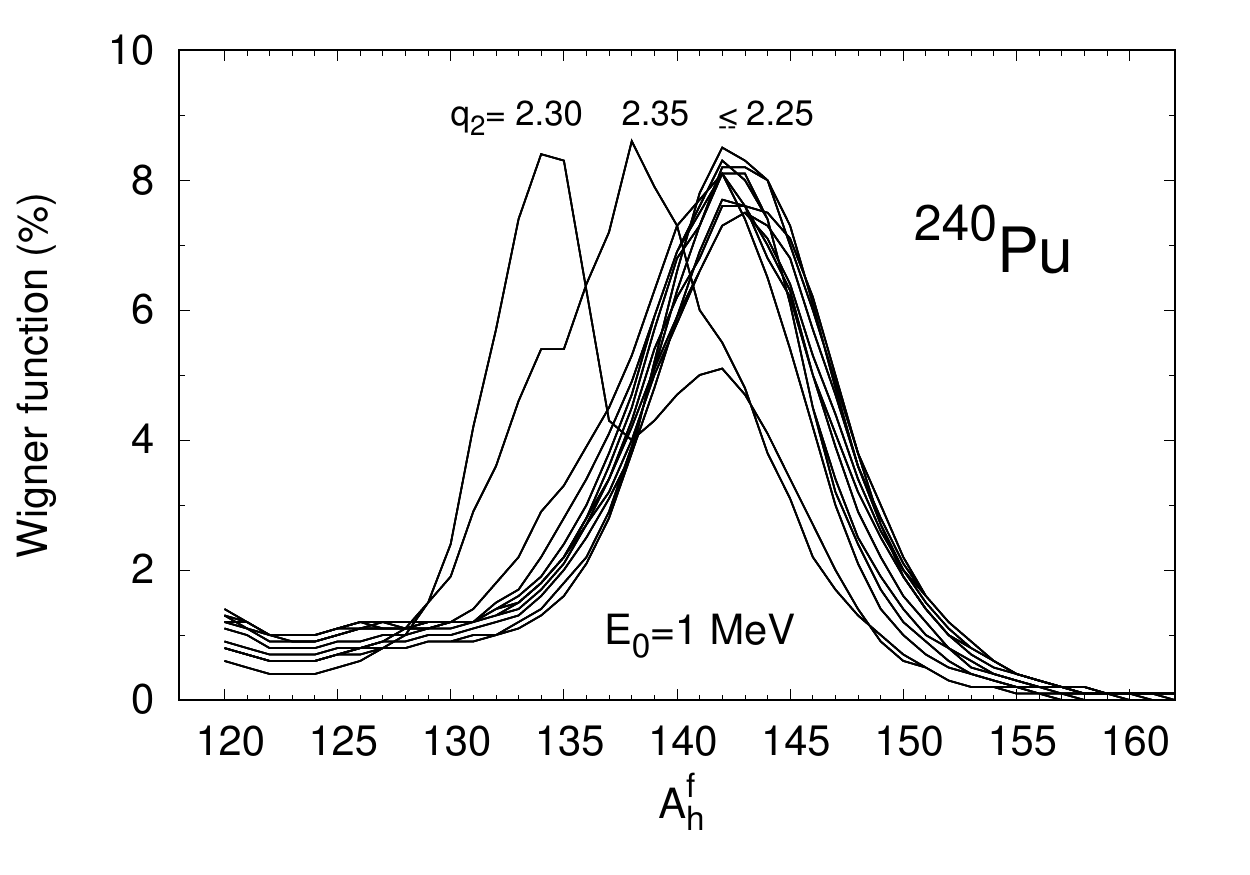}}
\vspace{-0.3cm}
\caption{Wigner function for different elongations $q_2$ as a function of the
heavy-fragment mass number $A^f_h$.} 
\vspace{-0.7cm}
\label{wigner}
\end{figure}

It is obvious that 2D maps do not give a complete information about the way in
which a system goes to fission. The neck dimension, which for a given
elongation $q_2$ and asymmetry $q_3$ depends on $q_4$, decides when fission
occurs. The map of the PES of $^{236}$Pu on the $(q_3,q_4)$ plane for $q_2=2.25$
is presented as an example in Fig.~\ref{e34}. Lines corresponding to a neck
radius equal to the one ($R_n$) of a nucleon, an alpha-particle ($R_\alpha$), as
well as of the heavier mass fragment $A^f_h=132$ and $A^f_h=140$ are marked in
the plot. It is seen that the deformation parameter $q_3$ governs mostly the
fragment mass, while $q_4$ is related to the neck size. The line corresponding
to the distance $R_{12}/R_0=2.25$ between the fragment mass centres is also
shown. Note that in all other points of the map $R_{12}$ differs from this mean
value by less than 5\% what proves that at large elongations $q_2$ is roughly a
measure of the distance between the fragments. 

More informative are the projections of the PES's for different elongations
$q_2$ onto the $(A^f_h,R_{\rm neck})$ maps presented in Fig.~\ref{erna} for
$^{236}$Pu. It is seen that the structure of the bottom of the fission valley
changes with increasing elongation of the nucleus. At smaller deformations
$q_2\le 2.25$, the bottom of the valley in the vicinity of $R_{\rm neck} = R_n$,
the size of a nucleon (thin solid), corresponds to the mass of the heavier
fragment $A^f_h=140$. At $q_2=2.3$ already two valleys at 138 and 144 are
visible, while at $q_2=2.35$ three fission valleys corresponding to 136, 141 and
146 are formed. The sizeable dependence of the PES in the  ($A^f_h, R_{\rm
neck}$) plane on the elongation coordinate $q_2$ shows how  important is to take
into account the subsequent steps in the fission process  as explained above.
\begin{figure}[h!]
\centerline{\includegraphics[width=0.6\textwidth,angle=0]{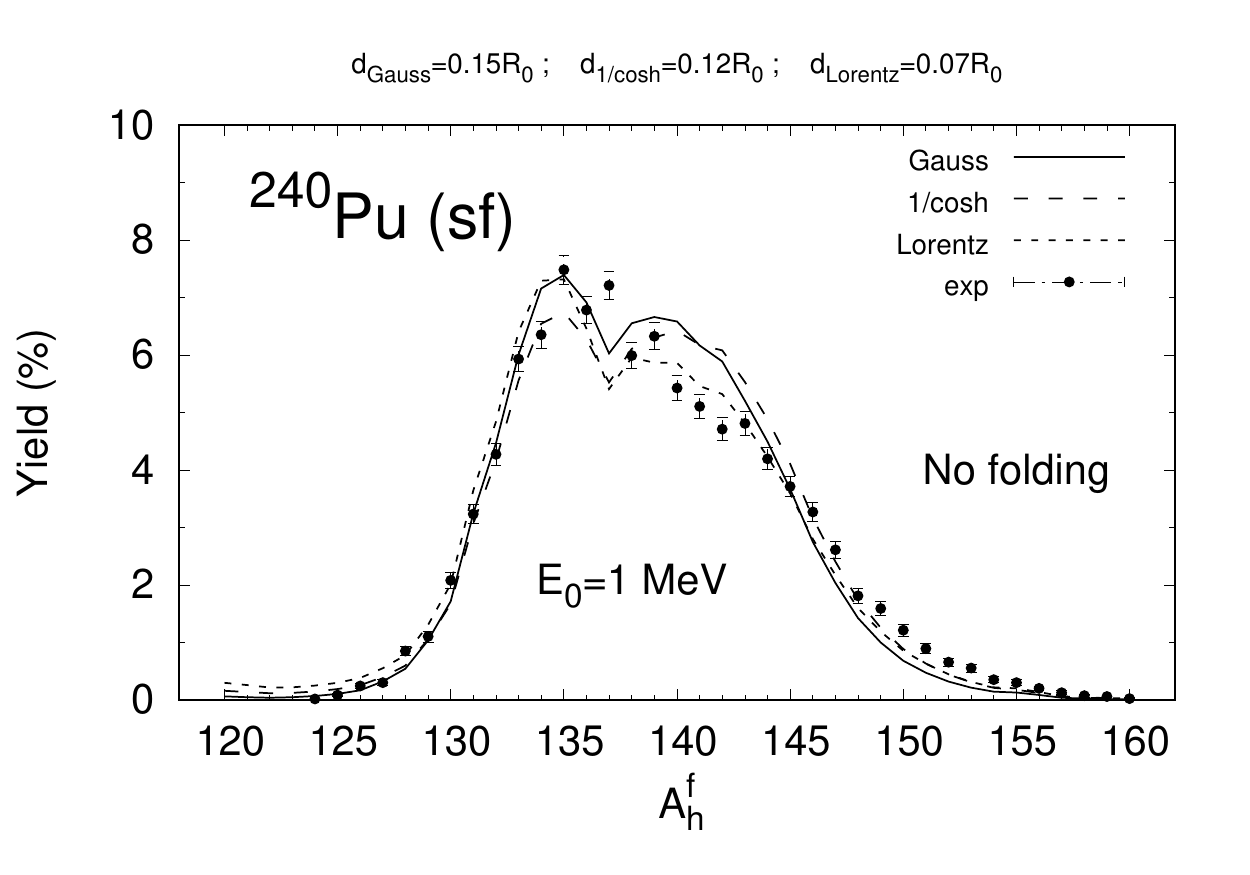}}
\vspace{-0.3cm}
\caption{Heavy fragment mass yield for fission of $^{240}$Pu as obtained with 
three different neck-breaking probability functions (\ref{pneck}).  The
half-width of each function is written above the plot. The calculation  is
compared to the yield measured in spontaneous fission in Ref.~\cite{Dematte}.} 
\vspace{-0.7cm}
\label{yield-1}
\end{figure}
\begin{figure*}[t!]
\centerline{
\includegraphics[width=0.3\textwidth,angle=0]{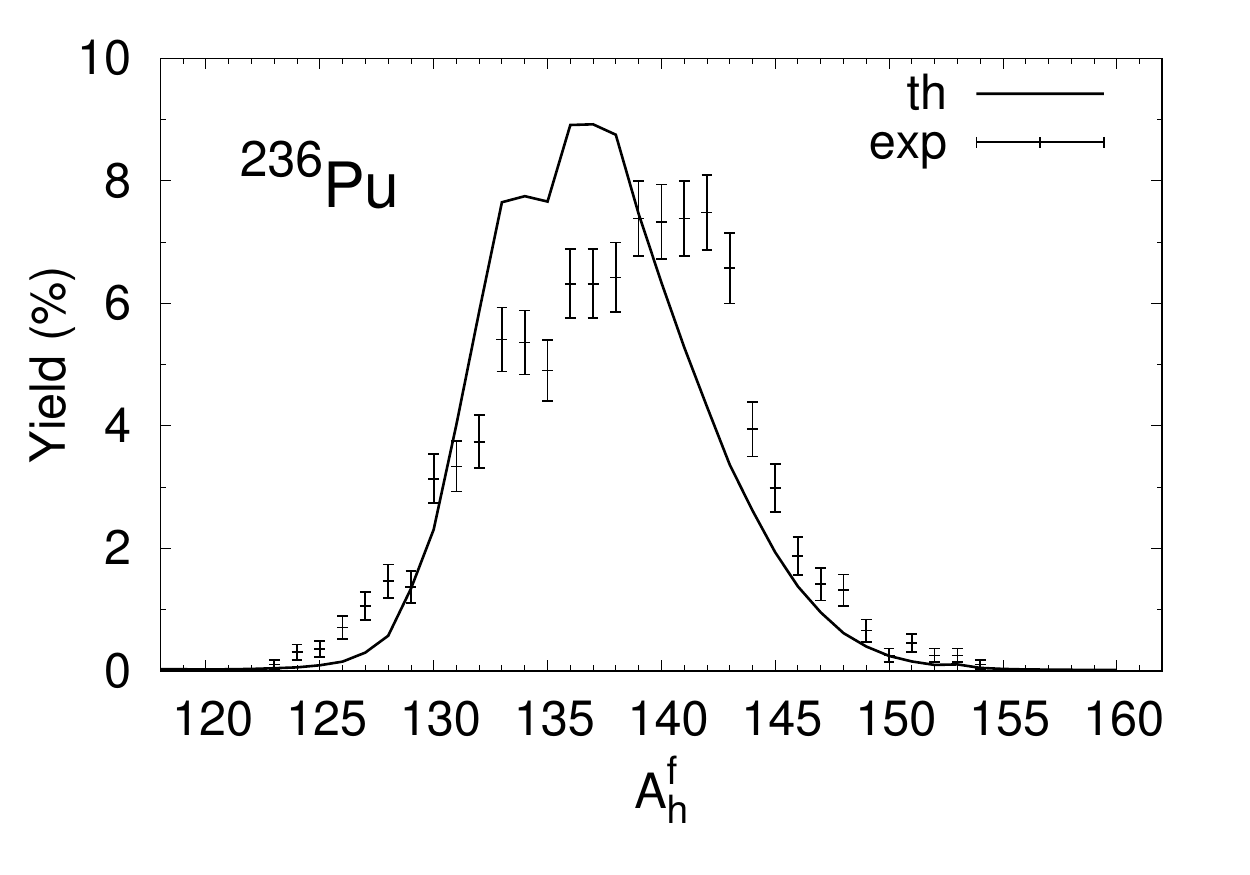}
\includegraphics[width=0.3\textwidth,angle=0]{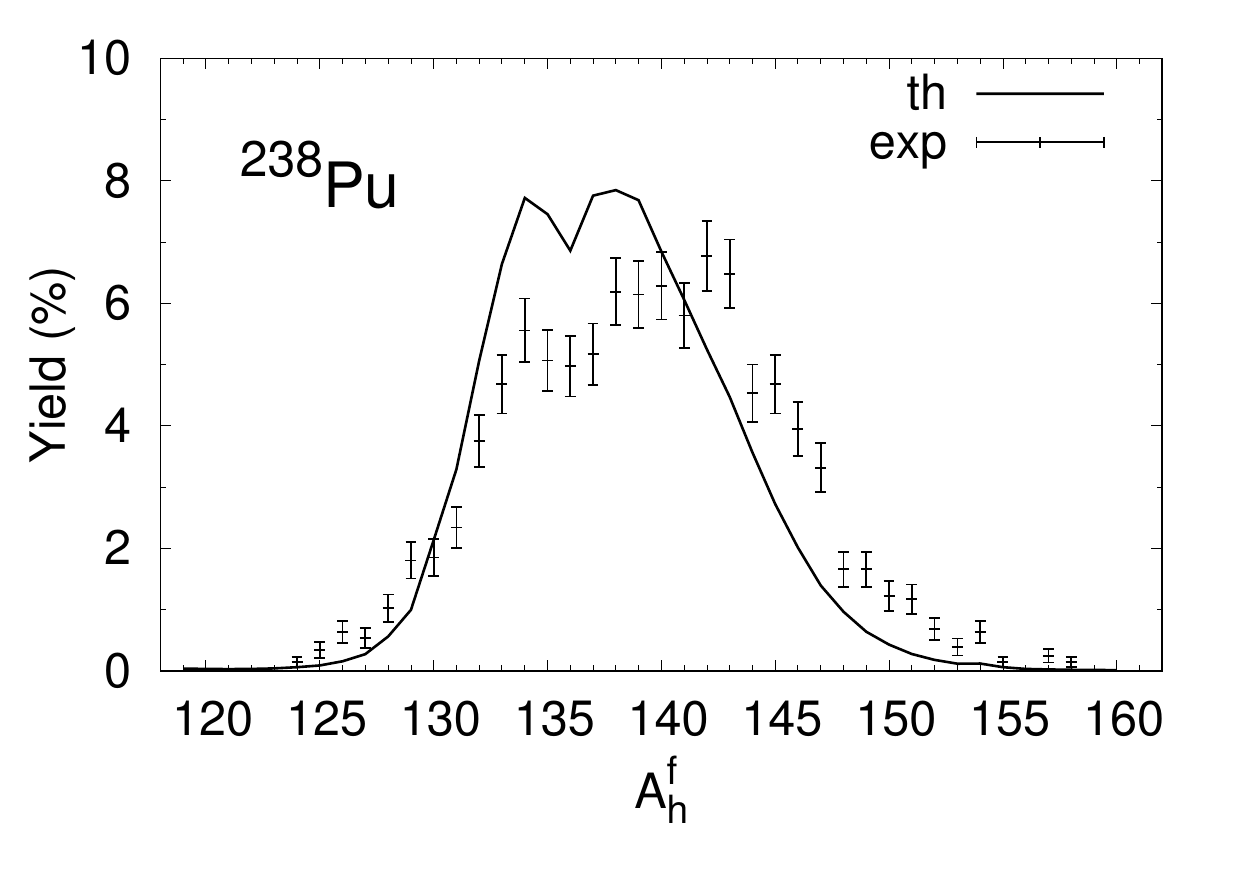}
\includegraphics[width=0.3\textwidth,angle=0]{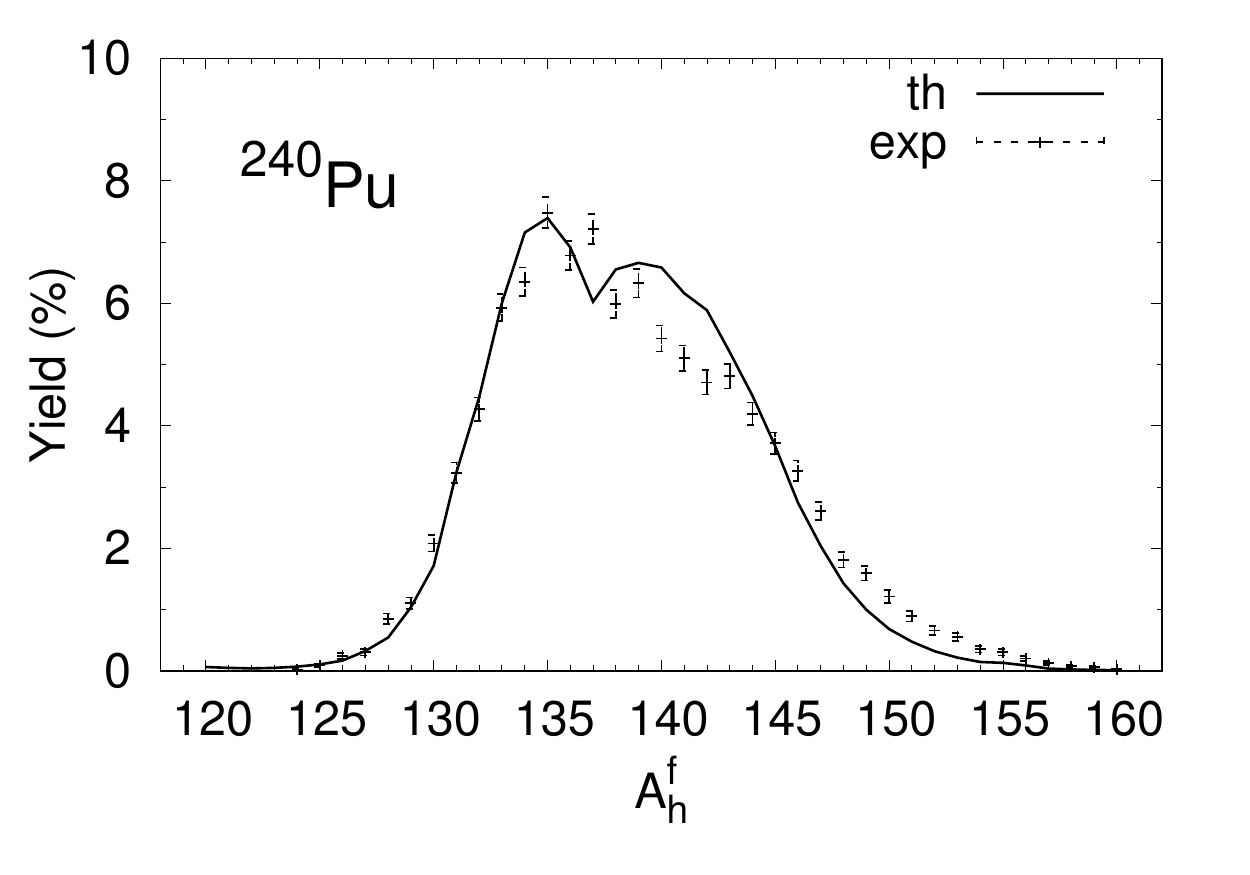}}
\centerline{
\includegraphics[width=0.3\textwidth,angle=0]{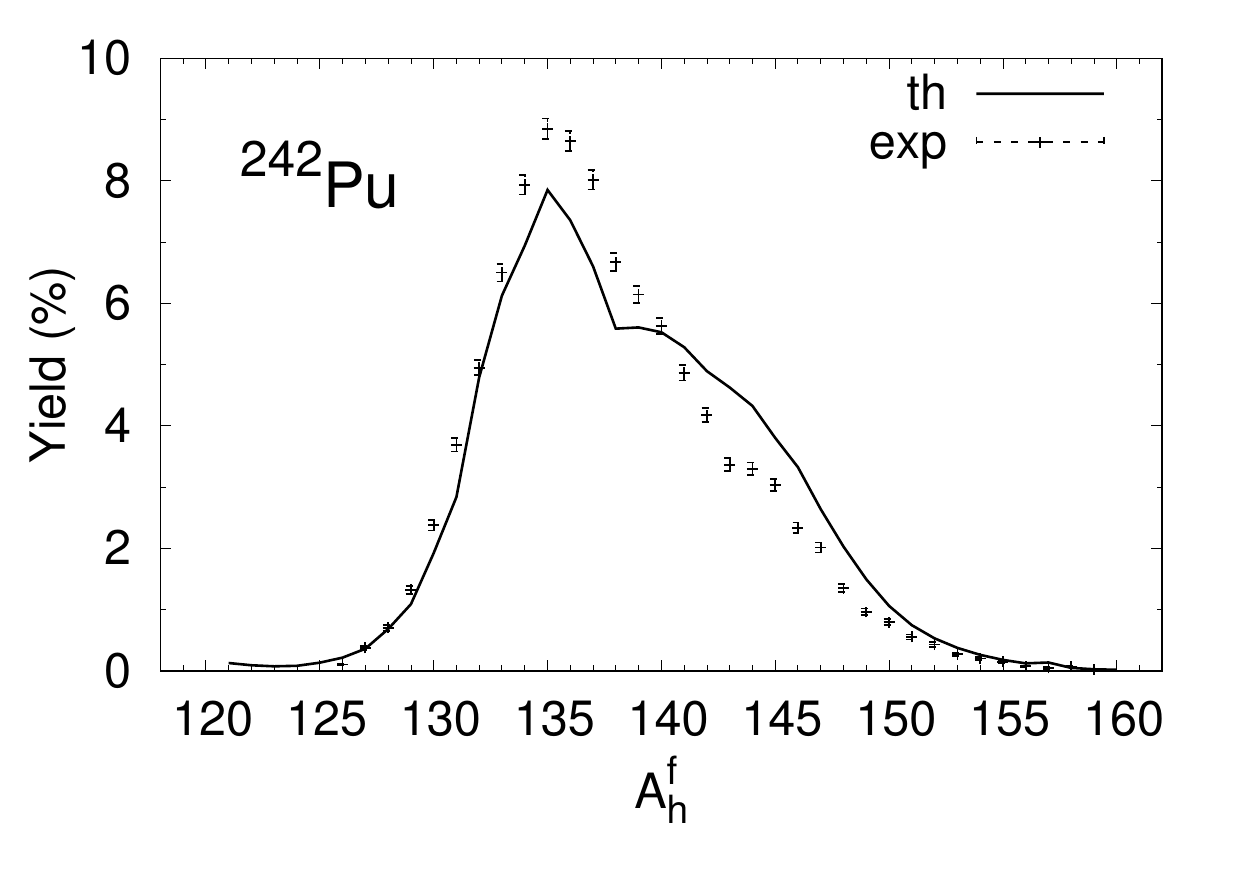}
\includegraphics[width=0.3\textwidth,angle=0]{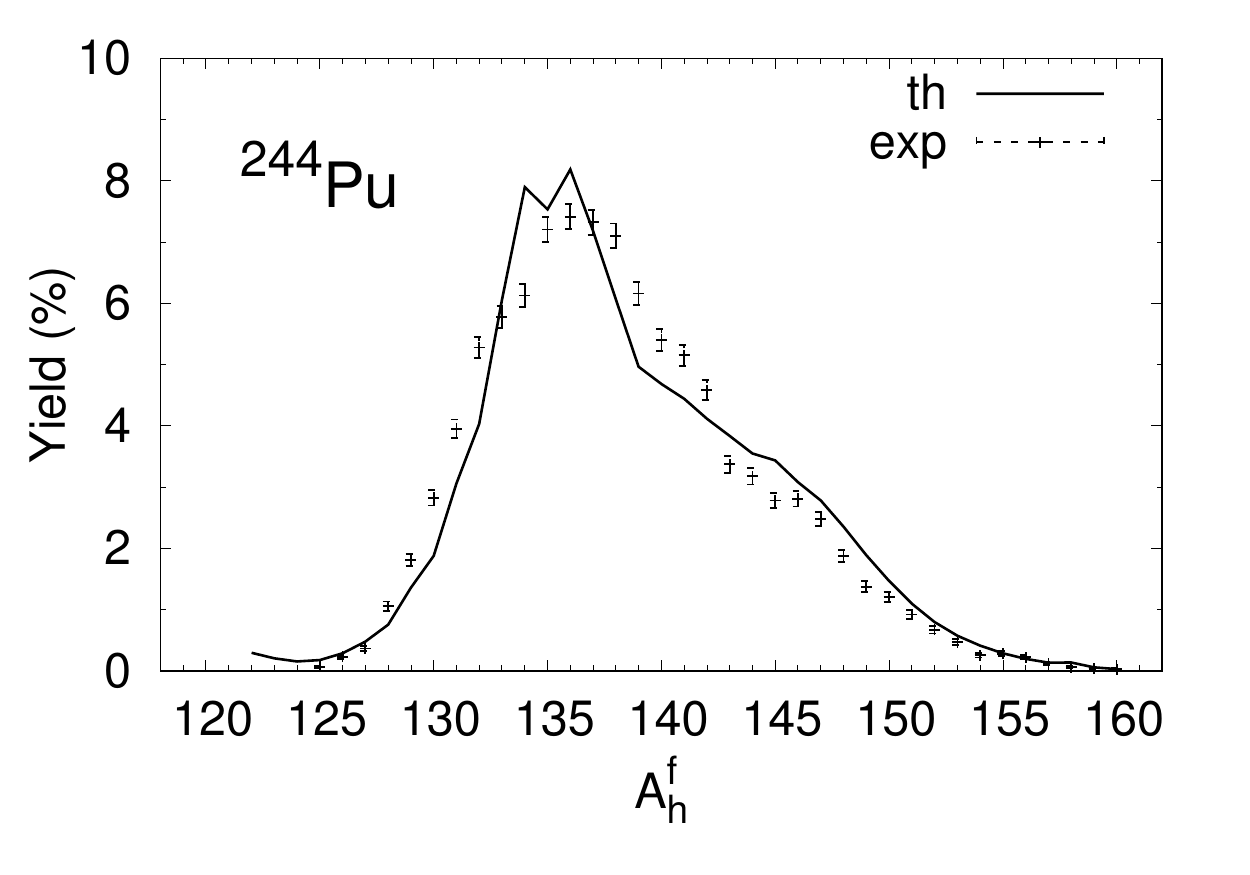}
\includegraphics[width=0.3\textwidth,angle=0]{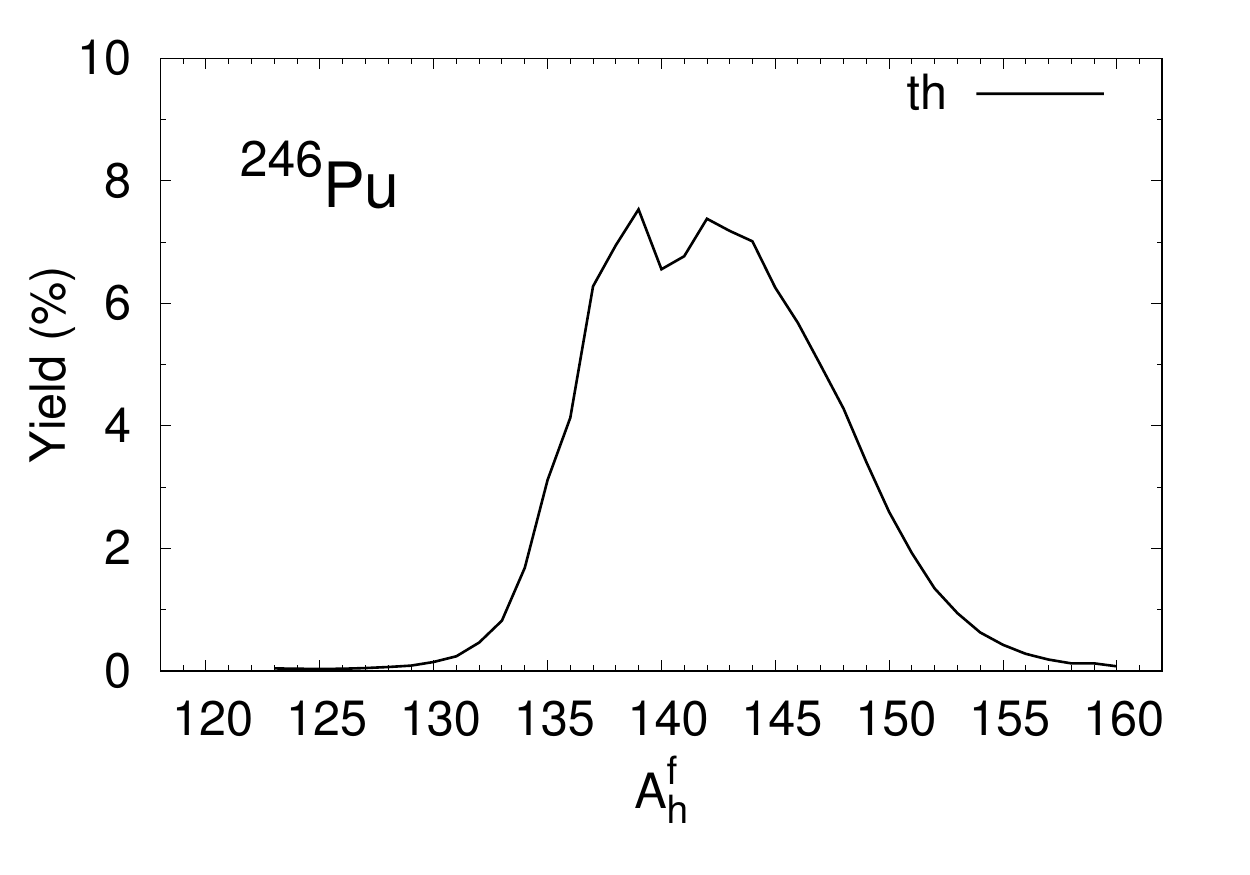}}
\caption{Same as in Fig.~\ref{yield-1} for fission of $^{236-246}$Pu but for 
the optimal calculation (see the text). The theoretical results, obtained here 
without any folding procedure, are compared to the yield measured in 
spontaneous fission in Ref.~\cite{Dematte}) for $^{236-244}$Pu.} 
\label{myields}
\end{figure*}

The Wigner function (\ref{waR}) integrated over $q_4$ is presented in
Fig.~\ref{wigner} for different elongations $q_2$ and projected onto the
$A^f_h$ axis, instead of the asymmetry parameter $q_3$. Its maximum changes
accordingly to the properties of the fission valleys shown in Fig.~\ref{erna}.
This set of Wigner functions weighted with the neck-size dependent fission
probability gives access to the mass yields (\ref{pda}) for different
elongations $q_2$. 

The final heavy-fragment mass yield (\ref{Ya}), obtained for three different
neck-breaking probabilities (\ref{pneck}) is shown in Fig.~\ref{yield-1} for
$^{240}$Pu, and compared to the yield measured in spontaneous fission
\cite{Dematte}. The differences between the theoretical yields are small as can
be seen in the figure. We shall therefore present in the following the results
obtained with the Gauss distribution only keeping the half-width fixed at
$d/R_0=0.15$ which roughly corresponds to the nucleon radius. The
width-parameter $E_0$ of the Wigner function was taken as $E_0$=1 MeV what
corresponds to a typical energy of collective modes in two dimensions ($q_3$ and
$q_4$). 

The spontaneous fission yields for six even-even Pu isotopes are compared in
Fig.~\ref{myields} with the experimental data whenever available \cite{Dematte}.
Please note that these theoretical estimates were obtained with only two
adjustable parameters $d/R_0=0.15$ and $E_0=1$ MeV. All other parameters of the
model are unchanged as compared to those adjusted years ago when considering
still different data.

\begin{figure}[thb]
\centerline{\includegraphics[width=0.5\textwidth,angle=0]{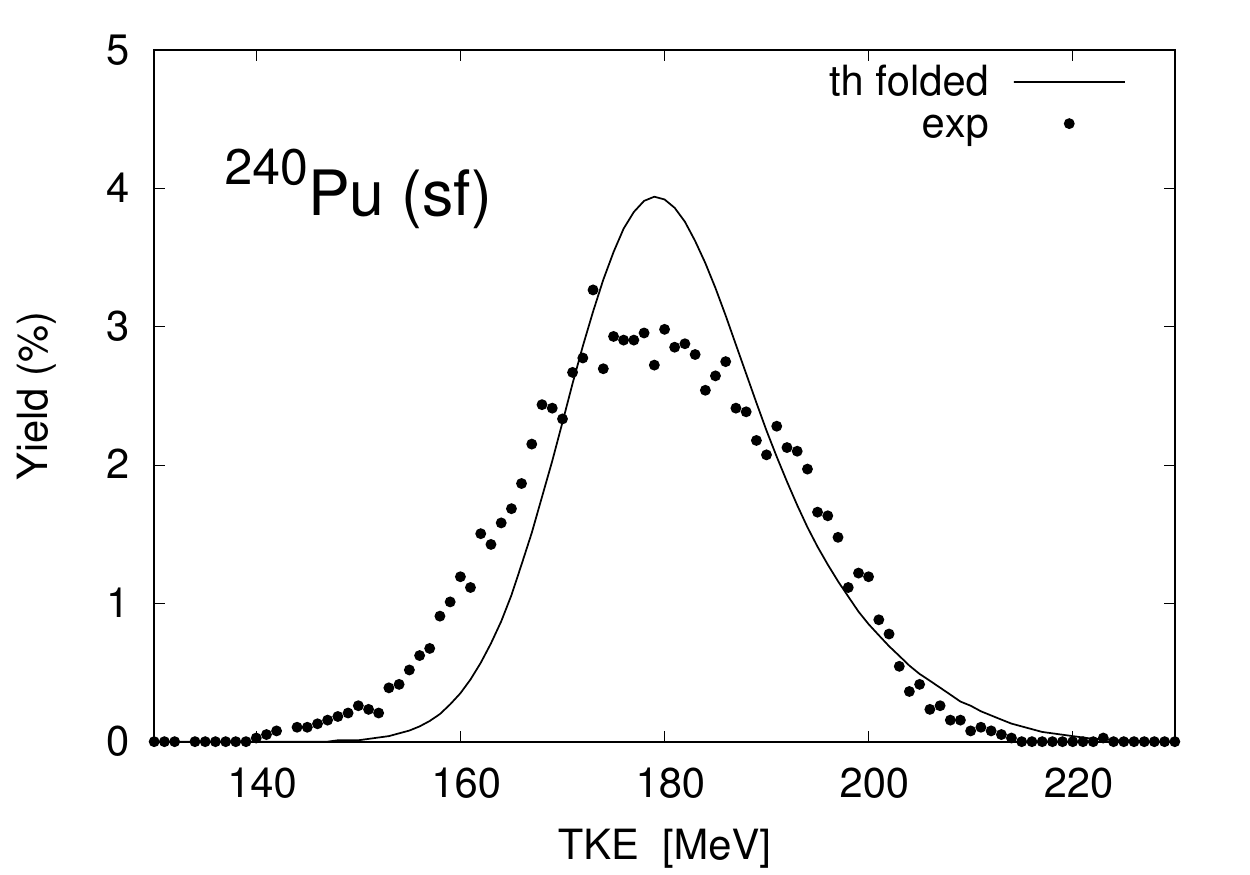}}
\centerline{\includegraphics[width=0.6\textwidth,angle=0]{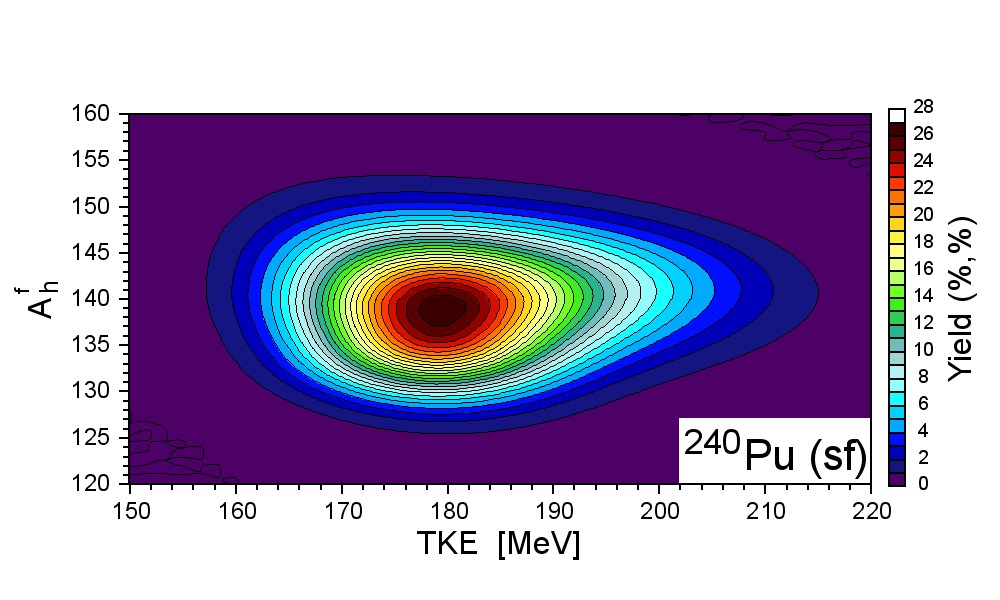}}
\caption{Top: Calculated and experimental (as measured in spontaneous-fission \cite{Wag}) TKE distribution for 
$^{240}$Pu. Bottom: Calculated correlation between the heavy-fragment mass and the TKE. Calculated mass and TKE values were folded with the experimental resolution (see text).} 
\label{yekamf}
\end{figure}

The total kinetic energy (TKE) of the fragments is evaluated in the point-charge
approximation of the Cou\-lomb interaction energy between the fragments. No
pre-fission kinetic energy is used in the present calculation, {\it i.e.} we
assume that the whole energy gain on the slope to scission is transferred into
heat. One has to bear in mind that for a given elongation $q_2$ the mass of the
fragment $A^f_h(q_3,q_4;q_2)$ and the TKE$(q_3,q_4;q_2)$ are functions of $q_3$
and $q_4$. So, it is easy to map the yields from the $(q_3,q_4)$ onto the
$({\rm TKE},A^f_h)$ plane. The experimental data have a finite resolution in mass and
in TKE, which in not negligible  for the present cases \cite{Dematte}. Typical
values are $\sigma_{\rm TKE}$=10 MeV  and a variance of $\sigma_{A^f_h}$=1.5 a.m.u.
Hence,  the theoretical $A^f_h$ and TKE yields have to be folded according to
these values before they can be compared  to the experiment. The resulting
folded TKE yield is presented in Fig.~\ref{yekamf} (top). One can see in the
figure that the  maximum of the TKE yield for $^{240}$Pu corresponds to TKE
$\approx$ 179 MeV what is in line with the measured data \cite{Wag}. The
calculated correlation  between the heavy-fragment mass and the TKE is presented
in Fig.~\ref{yekamf} (bottom). Although a quantitative analysis is still to be
done, it is already noted that the theoretical correlation shows a trend similar
to the  experiment \cite{OBER2017}, with a mean TKE larger for the  lighest
($A^f_h \approx 132$) heavy-fragment group, than for the heaviest ($A^f_h
\approx 140$) group.  The relatively good agreement of the TKE distribution and
mean value with the  experiment can be considered as quite encouraging. Keep in
mind, however, that  our theoretical approach is still quite crude, with the use
of a Wigner functions for the probability distribution instead of the true
quantum mechanical density distributions, the absence in our calculations of
extremely elongated shapes, since it was impossible to construct for them a
reliable Wigner function, the inclusion of only the lowest energy phonon in the 
direction perpendicular to the scission mode, i.e. neglecting the dynamical 
coupling of the fission and the perpendicular modes.  
%

\section{Summary}

Summing up the following conclusion can be drawn from our investigation:
\begin{itemize}
\item an effective three-dimensional set of Fourier deformation 
  parameters is used to describe the fission process,
\item a collective 3D model describing elongation, mass
  asymmetry and neck modes was introduced,
\item a Wigner distribution is assumed for the neck and
  mass asymmetry degrees of freedom,
\item a neck-breaking probability depending on the neck-size has been adopted,
\item our 3D model which couples fission, neck and
  mass asymmetry modes is able to describe the main features 
  of the fragment mass and total kinetic energy yields.
\end{itemize}

This contribution presents the present status of our research. Similar
calculations are going to be performed to describe low-energy fission of other
isotopes. In parallel, we aim to develop a Langevin code to study fission 
at higher energies with the new Fourier deformations space.
\section*{Acknowledgements:}
This work has been partly supported by the Polish-French COPIN-IN2P3
collaboration agreement project No.08-131 and by the Polish National
Science Centre, grant No. 2013/11/B/ST2/04087.
\section*{Appendix}
A new nuclear shape parametrization \cite{PNB15,PRC17} is used that gives an
expansion of the nuclear surface in the form of a Fourier analysis in
cylindrical coordinates
$$
\
 \frac{\rho_s^2(z)}{R_0^2} =\! \sum\limits_{n=1}^\infty \left[
 a^{}_{2n} \cos\left(\frac{(2n-1) \pi}{2} \, \frac{z-z_{\rm sh}}{z_0}\right)
 + a^{}_{2n+1} \sin\left(\frac{2 n \pi}{2} \, \frac{z-z_{\rm sh}}{z_0}\right)
\right] \; ,
$$
where, similarly to the famous ``Funny-Hills'' (FH) shape parametrization
\cite{FH72}, $\rho_s^{}(z)$ defines the distance of the equivalent sharp surface
from the symmetry $z$ axis, and $z_0$ is half the elongation of the nuclear
shape along that axis with end points located at $z^{}_{\rm min} \!=\! z_{\rm
sh} - z_0$ and $z^{}_{\rm max} \!=\! z_{\rm sh} + z_0$. The coordinate $z_{\rm
sh}$ is chosen so as to locate the centre of mass of the shape at the origin of
the coordinate system. $R_0$ represents the radius of the corresponding
spherical shape having the same volume and $z_0 \!=\! c \, R_0$.

The parameters $a_2, a_3, a_4$ are related to 
elongation, left-right asymmetry, and neck degree of freedom, respectively. 
More and more elongated prolate shapes correspond to decreasing values of $a_2$,
while oblate ones are described by $a_2$ larger than one, which is somehow
contrary to the traditional definition of a quadrupole deformation parameter.
So, it 
was proposed in Ref.~\cite{PRC17} to use a more {\it natural} definition
of deformation parameters which in addition ensures that along the LD fission
path these parameters show only a small variation around zero:
$$
\begin{array}{lll}
 q_2 = \frac{a_2^{(0)}}{a_2} - \frac{a_2}{a_2^{(0)}}~, &
 q_3 = a_3 \,, &           
 q_4 = a_4 + \sqrt{\left(\frac{q_2}{9}\right)^{\!2} +
   \left(a_4^{(0)}\right)^{\!2}} \,\,, \\
 q_5 = a_5 - (q_2-2) \, \frac{a_3}{10} \,\,,  & 
 q_6 = a_6 - \sqrt{\left(\frac{q_2}{100} \right)^{\!2} +
   \left(a_6^{(0)}\right)^{\!2}} \,\,. &
\end{array}
$$
Here $a_n^{(0)}$ stands for the value of the $a_n$
coefficient in the spherically symmetric case:\\
$a^{(0)}_2 = 1.03205 \,, ~~ a^{(0)}_4 =-0.03822 \,,~~
 a^{(0)}_6 = 0.00826 \,.
$

\end{document}